\documentclass{aa}  
\usepackage[draft]{hyperref}
\usepackage{amsmath}
\usepackage{graphicx}
\usepackage{amssymb}	
\usepackage{txfonts}
\usepackage[T1]{fontenc}

\begin{document} 

   \title{Taking apart the dynamical clock}
   \subtitle{Fat-tailed dynamical kicks shape the blue-straggler star bimodality}

   \author{Mario Pasquato
          \inst{1,2}
          \and
          Pierfrancesco Di Cintio
          \inst{3,4}
          }
   \institute{INAF, Osservatorio Astronomico di Padova, vicolo dell'Osservatorio 5, I--35122 Padova, Italy\\
              \email{mario.pasquato@oapd.inaf.it}
    \and
             INFN- Sezione di Padova, Via Marzolo 8, I--35131 Padova, Italy
    \and
             Dipartimento di Fisica e Astronomia \& CSDC, Universit\`a di Firenze, via Sansone 1, I--50019 Sesto Fiorentino, Italy\\
             \email{pierfrancesco.dicintio@unifi.it}
    \and
             INFN - Sezione di Firenze, via Sansone 1, I--50019 Sesto Fiorentino, Italy             
             }

   \date{Received April 12, 1633; accepted June 22, 1633}
  \abstract
   {In globular clusters (GCs), blue straggler stars (BSS) are heavier than the average star, so dynamical friction strongly affects them. The radial distribution of BSS, normalized to a reference population, appears bimodal in a fraction of Galactic GCs, with a density peak in the core, a prominent zone of avoidance at intermediate radii, and again higher density in the outskirts. The zone of avoidance appears to be located at larger radii the more relaxed the host cluster, acting as a sort of \emph{dynamical clock}.}
   {We use a new method to compute the evolution of the BSS radial distribution under dynamical friction and diffusion.}
   {We evolve our BSS in the mean cluster potential under dynamical friction plus a random fluctuating force, solving the Langevin equation with the Mannella quasi symplectic scheme. This amounts to a new simulation method which is much faster and simpler than direct N-body codes but retains their main feature: diffusion powered by strong, if infrequent, kicks.}
   {We compute the radial distribution of initially unsegregated BSS normalized to a reference population as a function of time. We trace the evolution of its minimum, corresponding to the zone of avoidance. We compare the evolution under kicks extracted from a Gaussian distribution to that obtained using a Holtsmark distribution. The latter is a fat tailed distribution which correctly models the effects of close gravitational encounters. We find that the zone of avoidance moves outwards over time, as expected based on observations, only when using the Holtsmark distribution. Thus the correct representation of near encounters is crucial to reproduce the dynamics of the system.}
   {We confirm and extend earlier results that showed how the dynamical clock indicator depends both on dynamical friction and effective diffusion powered by dynamical encounters. We demonstrated the high sensitivity of the clock to the details of the mechanism underlying diffusion, which may explain the difficulties in reproducing the motion of the zone of avoidance across different simulation methods.}
\maketitle
\section{Introduction}
Blue straggler stars (hereafter, BSS) were first observed by \cite{1953AJ.....58...61S} as a blueward and brighter continuation of the main sequence in the globular cluster (GC) M$3$, and have since been found in all GCs in the Milky Way \cite{2004ApJ...604L.109P,2020arXiv200107435F} and references therein.
BSS in dense stellar systems such as GCs have typical masses $m_{\rm BSS}$ of the order of twice the mean stellar mass $m_*$ in the host cluster and are born either through mass-transfer in close binary stars \cite{1964MNRAS.128..147M} or via direct stellar collisions \cite{1976ApL....17...87H}. More recently, it has been suggested that BSS may form also via hierarchical merging induced by Lidov-Kozai mechanism (\citealt{1962P&SS....9..719L,1962AJ.....67..591K}, see also \citealt{2007ApJ...669.1298F}) in triple systems (see e.g. \citealt{2009ApJ...697.1048P,2015ASSL..413.....B,Antonini_2016}).\\
\indent Each formation channel of BSS is favoured in different regions of the host GC. For example, the dense core allows for more collisions-induced BSS mergers \cite{1987IAUS..125..187V}. Early observational studies revealed a bimodal distribution of the BSS population in  GCs \cite{1993AJ....106.2324F, 1997A&A...327.1004Z}, which was interpreted as evidence that both channels are simultaneously active, and Monte Carlo simulations confirmed this interpretation \cite{2004ApJ...605L..29M, 2006MNRAS.373..361M}, as did scaling laws for the number of BSS with GC structural parameters \cite{2004MNRAS.349..129D}.
As later observational efforts increased the sample of GCs with a well-observed BSS radial distribution, including cases where no bimodality was present e.g. \cite{2006ApJ...638..433F, 2008ApJ...681..311D} it became apparent that the observed BSS bimodality is deeply linked to the dynamical relaxation of the host GC, to the point that it can be used as a sort of \emph{dynamical clock} to measure the evolutionary stage of a GC \cite{2012Natur.492..393F}. 
This finding, together with the proportionality of the BSS number with GC core mass e.g. \cite{2009Natur.457..288K} qualitatively supports a scenario where BSS originate from primordial binaries for the most part, as is also suggested by much more recent direct observations \cite{2019arXiv190904050G,Gosnell_2019}.\\
\indent However, a qualitative agreement is not enough to rule out the direct-collision channel (also supported by earlier hydrodynamical simulations by \citealt{1996ApJ...468..797L}), which could still bring a significant contribution to the BSS population in GC cores. In addition there have also been recent claims of observational evidence of possible ternary mergers as origin of BSS (\citealt{2016ApJ...828...38A,2018AAS...23124406K}) as well as evidences of BSS with white dwarfs companions (\citealt{2018MNRAS.479.2623E,2019arXiv190801573N}). As a matter of fact, a precise quantitative prediction of the mass-transfer BSS distribution as a function of time would allow us to obtain, by subtraction from the observations, the number of direct-collision BSS present in cores, if any exist.\\
\indent To be able to make this sort of clear cut quantitative predictions, a clear understanding of the physics underlying the dynamical clock is needed.
This is necessary for example to understand if the initial conditions of a simulation are located in the correct region of parameter space needed to obtain the observed formation and outward motion of the BSS distribution minimum, and whether the results of a small-scale simulation can be scaled up to describe a larger system. Realistic simulation studies based on direct N-body simulations \cite{2012Natur.492..393F, 2014ApJ...795..169A, 2015ApJ...799...44M, 2016ApJ...833..252A, 2016MmSAI..87..513A} or state-of-the-art Monte Carlo simulations \cite{2013MNRAS.429.1221H, 2017MNRAS.471.2537H, 2019MNRAS.483.1523S}, while valuable for a direct comparison with observations in the spirit of \emph{saving the phenomena}, are less concerned with gaining this kind of understanding.
In a previous paper, \cite{2018ApJ...867..163P} made some progress towards this goal by showing that:
\begin{itemize}
    \item to form a minimum, i.e. to obtain a bimodal BSS distribution, dynamical friction is a necessary ingredient
    \item to move the minimum to larger radii over time, an effective diffusion mechanism is needed
    \item to obtain a bimodal distribution with a minimum that moves outwards over time, dynamical friction and diffusion should be balanced within a narrow range.
\end{itemize}
The latter condition in particular is not necessarily trivial to achieve within a simulation, but in real systems it stems naturally from fundamental fluctuation-dissipation relations that connect dynamical friction and diffusion, which are ultimately two aspects of the same phenomenon (\citealt{1980PhR....63....1K,1981Ap&SS..80..443K}).\\
\indent \cite{2018ApJ...867..163P} was based on a one dimensional Brownian motion model, with particles representing the average radial positions of stars over their orbits. The limitations of this model coincide with the limitations of the intuitive picture of BSS sitting undisturbed at large radii until the \emph{zone of avoidance} \cite{2004ApJ...605L..29M} reaches the scale radius of their orbit as dynamical relaxation takes place: namely, that in a three-dimensional system with non-circular orbits the radial position of stars changes over a timescale much shorter than the systems relaxation time simply because over an orbital period they move from the apocenter to the pericenter of their orbit and back. Incidentally, this was mentioned by \cite{2017MNRAS.471.2537H} as the probable reason due to which the external regions of a simulated GC suffer a quick depletion of BSS even though the zone of avoidance has not yet reached them (see e.g. their Fig. 4).\\
\indent In this work we solve a full three dimensional model with stochastic differential equations for a population of non-interacting tracer particles representing the BSS evolving under the combined effect of the cluster potential, dynamical friction and collisions with other stars. Such method allows one to run many BSS trajectories without interfering with the cluster dynamics (as they get kicks from the cluster stars, but they do not give any feedback to them), so that better statistics can be obtained, while avoiding to trigger \cite{1969ApJ...158L.139S} instability.\\
\begin{figure}
    \centering
    \includegraphics[width=0.96\columnwidth]{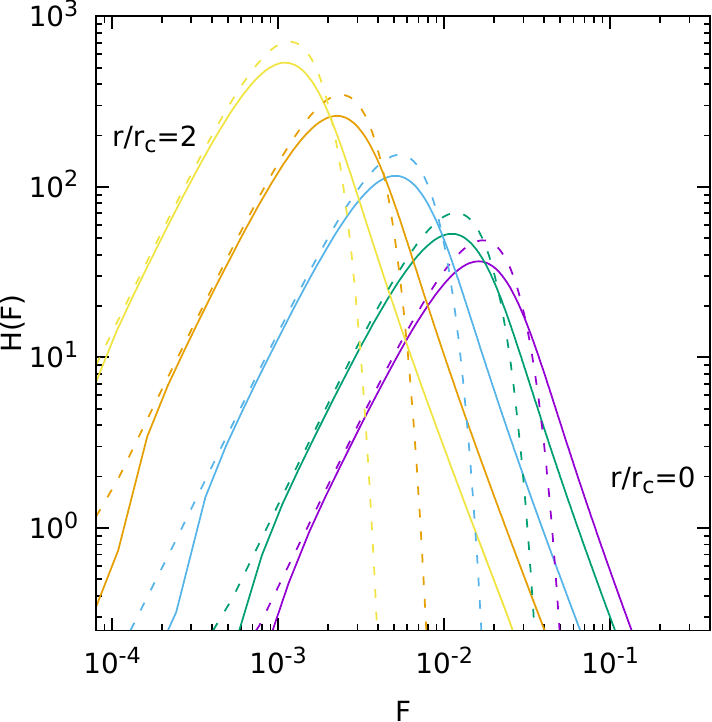}
    \caption{Holtsmark (solid lines) and Maxwellian (dashed lines) distributions of the intensity of the gravitational force fluctuations at $r/r_c=0$ (purple), 0.5 (green), 1 (light blue), 1.5 (orange) and 2 (yellow) in a Plummer model.}
    \label{disth}
\end{figure}
\indent Moreover, at variance with other stochastic schemes based on the evolution of probability density functions (PDFs) for the particles phase-space coordinates, with the Fokker-Planck equation (see \citealt{1957PhRv..107....1R} for the specific case of the $1/r^2$ force), frequently used to investigate the dynamics of black holes under the effect of gravitational encounters with stars (for example, see \citealt{1978ApJ...226.1087C,2002ApJ...572..371C,2015ApJ...804...52M,2015ApJ...804..128M,2019MNRAS.482.2132D}), our approach can be easily tuned to include for example, but not only, a mass spectrum (\citealt{2010AIPC.1242..117C}), orbital anisotropy or the effect of a time-dependent external potential.\\
\indent This paper is structured as follows: in Section \ref{modmethod} we present the numerical scheme to solve the Langevin equations for the population of BSS under the effect of the cluster potential and stellar collision. In Section \ref{results} we present the results of our numerical calculations and discuss them in light of the observational results. Finally, Section \ref{discussion} summarizes.
\section{Methods}\label{modmethod}
\label{sec:sims} 
We evolve an ensemble of non-interacting tracer particles (simulating the population of BSS) of equal mass $m_{\rm BSS}$, under the combined effect of the {\it fixed} star cluster mean field potential and  stellar encounters resulting in a diffusion and friction process. Under these assumptions, the particle dynamics is described by the Langevin equation (see e.g. \citealt{1992sppc.book.....V})
\begin{equation}\label{langeq}
\ddot{\mathbf{r}}=-\nabla\Phi(\mathbf{r})-\eta\mathbf{v}+\mathbf{F}(\mathbf{r});\quad \mathbf{v}=\dot{\mathbf{r}},
\end{equation}
where $\nabla\Phi$ is a smooth deterministic force field generated by the chosen spherical mass density, $\eta$ is the dynamical friction coefficient (\citealt{1943ApJ....97..255C,1949RvMP...21..383C}), and $\mathbf{F}$ is a {\it fluctuating} force per unit mass, accounting for the ``granular" nature of the underlying model. A similar approach has been used to treat different problems involving noise induced phase-space transport in the contexts of galactic dynamics (\citealt{1997ApJ...480..155H,1999PhRvE..60.1567P,2000MNRAS.311..719K,2003astro.ph.12434T,2004ApJ...602..678S}) and charged particle beams (\citealt{2004PhRvS...7j4202S,2004PhRvS...7a4202K}). More recently, \cite{modest19} used this method to study the dynamics of the black holes at the center of elliptical galaxies or star clusters, finding in the latter case a good agreement with simple direct $N-$body simulations.\\
\indent In Eq.~(\ref{langeq}) the fixed gravitational potential $\Phi$ is generated by the usual \cite{1911MNRAS..71..460P} density profile\footnote{The choice of the simple Plummer model, following \cite{2015ApJ...799...44M}, is motivated mainly by the fact that it possesses analytic and relatively simple expressions for the potential and velocity dispersion, at variance with the more realistic \cite{1966AJ.....71...64K} profile.}
\begin{equation}\label{plummer}
\rho(r)=\frac{3}{4\pi}\frac{Mr_c^2}{(r_c^2+r^2)^{5/2}},
\end{equation}
with total mass $M$ and scale radius $r_c$. The position-dependent dynamical friction coefficient (\citealt{2015ApJ...799...44M}) is defined as
\begin{equation}\label{eta}
\eta(r,v)=4\pi G^2m_*(m_{\rm BSS}+m_*)\ln\Lambda\frac{\Psi(r,v)}{v^3},
\end{equation}
where $v=||\mathbf{v}||$, $\ln\Lambda$ the Coulomb logarithm of the maximum to minimum impact parameter ratio $b_{\rm max}/b_{\rm min}$, and
\begin{equation}\label{velfunct}
\Psi(r,v)=4\pi\int_0^{v}f(r,v^\prime)v^{\prime 2}{\rm d}v^\prime
\end{equation}
is the so-called velocity volume function. In the case of a Plummer model, the isotropic phase-space distribution function is written simply as $f(r,v)=C[-\Phi(r)-v^2/2]^{7/2}$, where $C$ is the normalization constant and $\Phi(r)=-GM/\sqrt{r_c^2+r^2}$.\\
\indent At variance with the one-dimensional `orbit gas' model of \cite{2018ApJ...867..163P}, our model features isotropic kicks in three-dimensions and we sample the norm $F$ of the stochastic acceleration term in Equation (\ref{langeq}) from the \cite{1919AnP...363..577H} distribution
\begin{equation}\label{holtsmark}
H(F)=\frac{2}{\pi F}\int_0^\infty\exp\left[-\alpha(\xi/F)^{3/2}\right]\xi\sin(\xi){\rm d}\xi,
\end{equation}
introduced originally in the context of plasma physics, and used for the first time in stellar dynamics by \cite{1942ApJ....95..489C,1943ApJ....97....1C} to study the fluctuations of the gravitational field acting on a test star. In Equation (\ref{holtsmark}) $\alpha=(4/15)(2\pi G m)^{3/2}n$ is a normalization factor dependent on the number density $n$ and stellar mass $m$. Note that, in the original derivation by Chandrasekhar and von Neumann, Equation (\ref{holtsmark}) is defined for an infinite and homogeneous system. In this work we assume a position-dependent Holtsmark distribution by substituting $n_*(r)=\rho(r)/m_*$, i.e. the {\it local} mean number density, in the normalization parameter $\alpha$. However, (see e.g. \citealt{1958ApJ...128..130B}) for sufficiently flat-cored models the distribution of force fluctuations differs little from the Holtsmark distribution.\\
\indent Unfortunately, it is not possible to write the Equation (\ref{holtsmark}) and its cumulative distribution in terms of simple functions. Moreover, except the first, the moments of the distribution are all singular\footnote{We note that, (see \citealt{1986SvAL...12..237P}) an approximated expression for the Holtsmark distribution with finite normalization and standard deviation can be obtained by substituting Eq. (\ref{holtsmark}) for $F\geq F_*$ with $H_1(F)=(\alpha^{4/3} L/F^3)\exp(-3F^2/2 \alpha^{4/3})$, where $L\approx5.2\sqrt{6\pi}n_*b_{\rm min}^3$ and $F_*$ is tuned so that the two expression mach for $F=F_*$.}, which makes sampling the stochastic force term in Equation (\ref{langeq}) a delicate step.\\
\indent In the limit of large $F$, the Holtsmark distribution (\ref{holtsmark}) is well approximated in polynomial form (see \citealt{1986JQSRT..36....1H}) and, retaining only the leading term of the expansion, it can be written as $\tilde H(F)\sim 2\pi n_*(Gm_*)^{3/2}F^{-5/2}$. The latter expression, frequently used in numerical studies in the cosmological context (see \citealt{2002JPCM...14.2141P,2002EL.....57..315B}, and references therein) still bears the same problems of its full integral form, being non-normalizable and with divergent standard deviation. Typically, in order to avoid a diverging cumulative distribution and diverging energy density of the fluctuating field (\citealt{Kozlitin2011}), when sampling $\tilde H(F)$ in numerical schemes one is forced to fix bona fide cut-offs at large and and small $F$.\\
\indent In the numerical simulations discussed in this work we have used the integral representation of the Holtsmark distribution whose cumulative function   
\begin{equation}\label{cumulative}
    C(F)=\int_0^FH(F^\prime)F^\prime{\rm d}F^\prime
\end{equation}
has been evaluated numerically on a unevenly spaced grid between 0 and and a maximal force of the order of $Gm_*/\beta^2$, where $\beta$ is the {\it typical} minimum impact parameter, which we set at $1/30$ of the local mean inter-particle distance, corresponding to roughly the size of the Solar system at $r\approx 2 r_c$ for a star cluster of $10^6$ stars with a scale radius $r_c$ of $1$ pc.\\
\indent In Fig. \ref{disth} we show the numerically recovered Holtsmark distribution of the intensity of the force fluctuations at different radii $0\leq r\leq 2r_c$ in a Plummer model, and the associated Maxwell-Boltzmann distributions peaking at the same $F$, resulting from assuming a 3D Gaussian distribution of force fluctuations. As one would expect, the peaks of the Holtsmark and Gaussian distributions both drift towards lower forces as the radius increases. However, at all radii the Gaussian underestimates the contribution of strong kicks, corresponding to small impact parameters encounters between stars, with respect to the parent Holtsmark distribution.\\
\indent Equation (\ref{langeq}) is an example of stochastic ordinary differential equation (e.g., see \citealt{1994hsmp.book.....G}) for a single ``Brownian particle\footnote{Technically speaking one has Brownian motion under the assumptions that the stochastic force is isotropic, delta-correlated and normally distributed. The latter assumption is invalid in our case, as we are considering a fluctuating force described by the Holtsmark distribution.}" whose integration in general presents several technical issues due to the fluctuating nature of the stochastic force term $\mathbf{F}(\mathbf{r})$ 
(see e.g. \citealt{SanMiguel2000,doi:10.1137/050646032}, and references therein). In this work we use the so-called quasi-symplectic method, introduced in \cite{2004PhRvE..69d1107M}, that for the one dimensional case reads
\begin{eqnarray}\label{mannella}
x(t+\Delta t/2)=x(t)+\frac{\Delta t}{2}v(t)\nonumber\\
v(t+\Delta t)=c_2\left[c_1v(t)+\Delta t \nabla\Phi(x^\prime)+d_1 \tilde F(x^\prime) \right]\nonumber\\
x(t+\Delta t)=x(t+\Delta t/2)+\frac{\Delta t}{2}v(t+\Delta t).
\end{eqnarray}
In the equations above $\Delta t$ is the fixed time-step (we usually take $\Delta t\sim 10^{-3}t_c$, with $t_c\equiv\sqrt{r_c^3/GM}$ the crossing time of the system),  $\tilde F$ is the {\it normalized} stochastic force (in this case, a random variable sampled from Eq. \ref{holtsmark}), and
\begin{equation}
c_1=1-\frac{\eta\Delta t}{2};\quad c_2=\frac{1}{1+\eta\Delta t/2};\quad d_1=\sqrt{2\zeta\eta\Delta t},
\end{equation}
where $\zeta$ in the case of a delta correlated noise is fixed by the standard deviation of the distribution of $F$ as
\begin{equation}\label{sd}
\langle F(x,t) F(x,t^\prime)\rangle=2\eta\zeta\delta(t-t^\prime).
\end{equation}
\begin{figure*}
    \centering
    \includegraphics[width=0.9\textwidth]{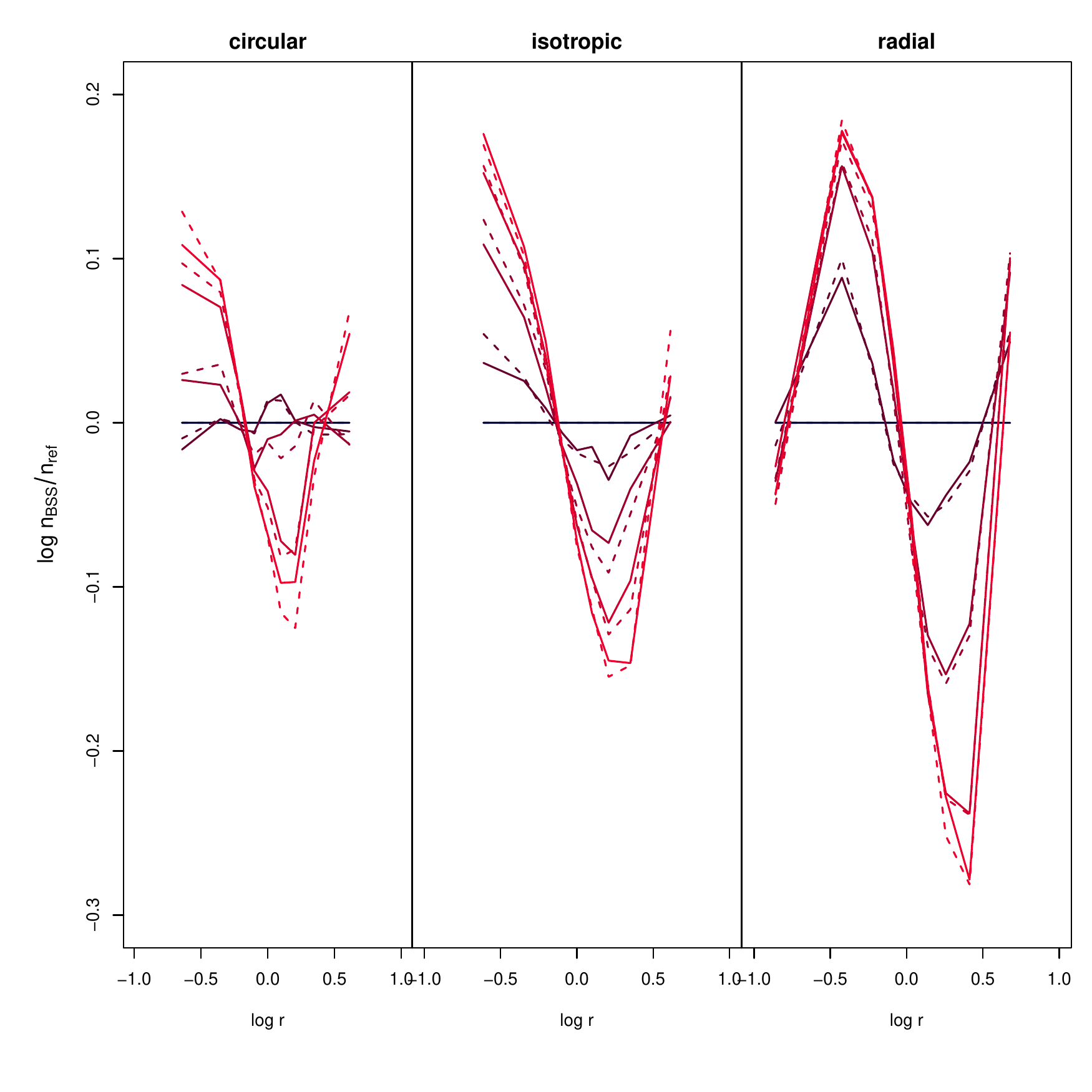}
    \caption{Normalized BSS distribution $n_{\rm BSS}/n_{\rm ref}$ at $t/t_c=0$ (black), $500$, $5000$, $20000$, $50000$, and $80000$ (bright red) for a population of BSS with masses $m_{BSS}=1.5m_*$ (dashed lines) and $m_{BSS}=2m_*$ (solid lines), initially placed on purely circular orbits (left), extracted from a isotropic distribution (center), and on purely circular orbits (right).}
    \label{holt}
\end{figure*}
Since for the Holtsmark distribution, the standard deviation and all higher moments are infinite, in our numerical scheme we use $1/\sqrt{8\log{2}} \approx 0.425$ of the full width at half maximum of the truncated distribution\footnote{For a normal distribution FWHM $= \sigma\sqrt{8\log{2}}$.} in place of $\sigma$. We note that (see also Fig. \ref{disth}), given that the two distributions differ of several orders of magnitude for large $F$, the results are left unchanged for other choices of the range $1/2,3/2$ of the full width at half maximum. Note also that, for vanishing $\eta$ and $\zeta$, Eqs. (\ref{mannella}) yield back the standard Leapfrog method, that is second order and symplectic. Generalization to higher order scheme is also possible see \cite{doi:10.1137/040620965,doi:10.1137/050646032};
however, for the scope of this paper we limited ourselves to the second order method. Using a second order method to solve Eq. (\ref{langeq}) allows to keep relatively small computational times even for a large number of test particles while avoiding to apply the contribution of the random force and dynamical friction {\it a posteriori} after a propagation step in the same fashion as \cite{1995ApJS...99..609S} for the interaction of Neutron stars with binaries in GCs. 
\begin{figure*}
    \centering
    \includegraphics[width=0.9\textwidth]{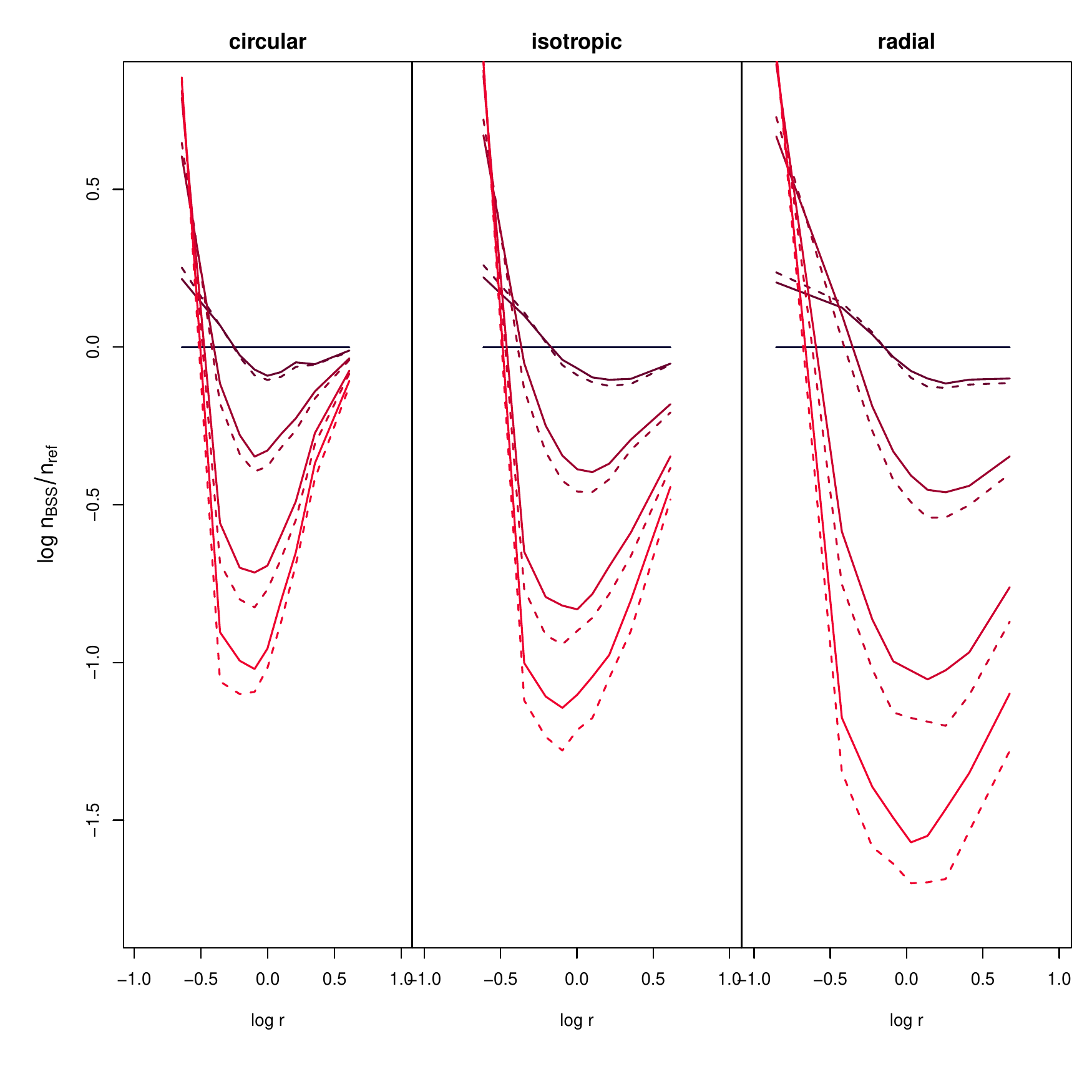}
    \caption{Same as in Fig. \ref{holt}, but for a Gaussian distributed force fluctuations field.}
    \label{gauss}
\end{figure*}
\subsection{Models}
In this study we have evolved $10^4$ independent\footnote{Particles representing the BSS are propagated separately according to Eq. (\ref{langeq}) and their distribution does not affect the fixed GC potential.} BSS with two values of mass $m_{\rm BSS}=1.5m_*$ and $2m_*$, for $10^5t_c$. The background population of the GC is assumed in both cases to be $10^6$ stars with equal masses $m_*$.\\
\indent The initial positions and velocities of the BSS are also drawn from the isotropic  Plummer distribution. In order to evaluate the effects on the mass segregation of radial or tangential anisotropy, we have also performed numerical integrations with the same parameters and initial positions of the BSS but sampling their velocities from the two extreme cases of a purely radial or circular orbit distribution, as done for example in the case of pulsar binaries (\citealt{1991Natur.349..220P,1995ApJS...99..609S}). In addition, to evaluate the importance of correctly modeling close encounters (represented by the fat tails of the Holstsmark distribution), we have performed an additional set of numerical experiments using for the diffusion term in Eq. (\ref{langeq}) a 3D isotropic Gaussian distribution of force fluctuations (corresponding to a Maxwell-Boltzmann distribution of their modulus), instead of the Holtsmark distribution.\\
\indent In each simulation set-up we have extracted the projected distribution of the $n_{\rm BSS}$ population normalized to the reference background population that was assumed to remain constant in time.\\
\indent In order to build the radial profile at a given time $t_p$, we took the eight subsequent snapshots (taken one dynamical time apart from each other) and we merged them into one then pooling together all the stars projected along each of the three coordinate axes. This allows us to virtually increase the number of BSS in the sample by a factor $24$, further reducing statistical fluctuations. Note that, merging subsequent snapshots is justified by the fact that we are interested in the long-term dynamical evolution of the system, with respect to which snapshots separated by one dynamical time are essentially identical. The projected profiles were then binned in ten radial bins based on the quantiles of the Plummer density distribution, so that an equal number of reference stars would fall in each bin.
\section{Results}\label{results}
\label{simulations}
\begin{figure}
    \centering
    \includegraphics[width=0.9\columnwidth]{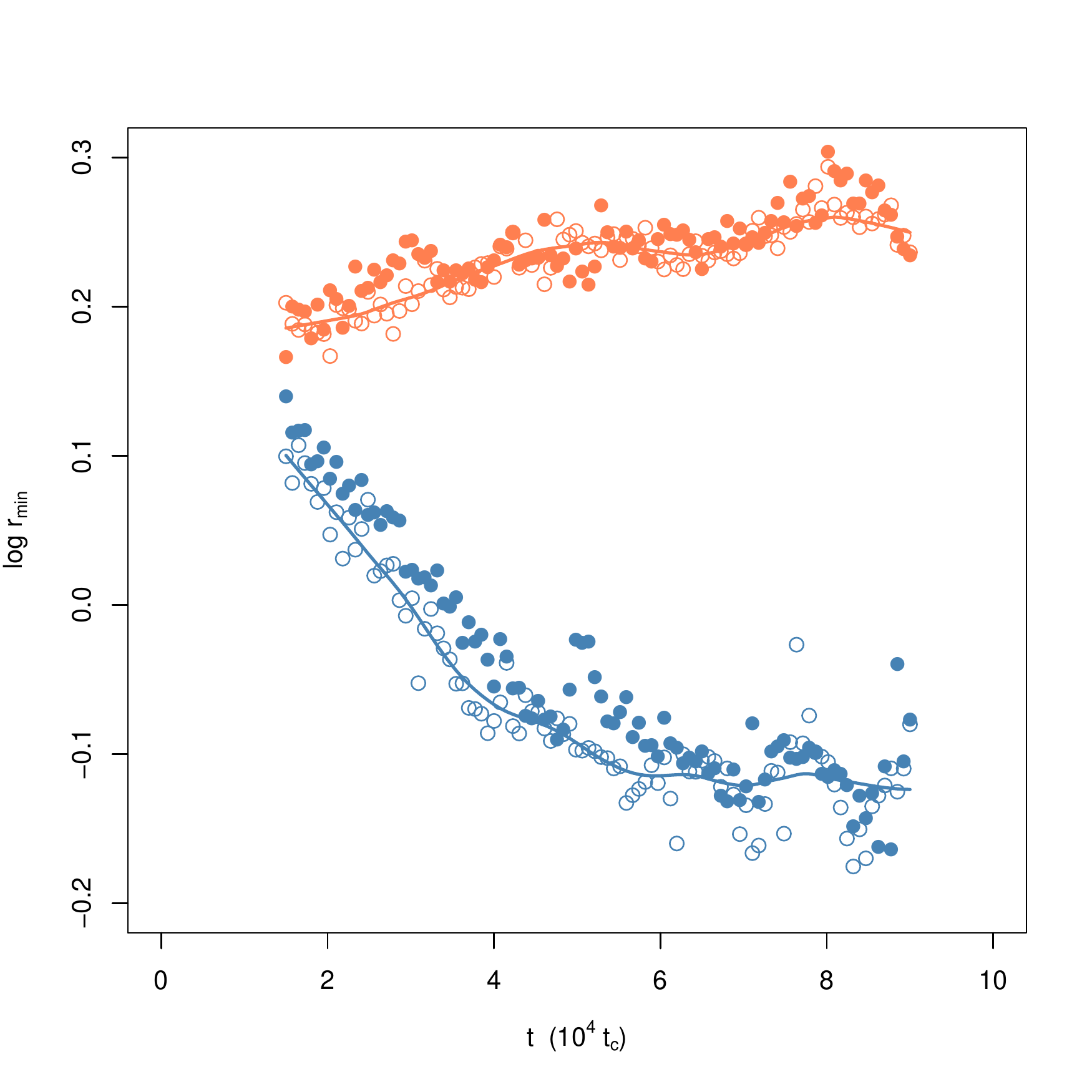}
    \caption{Time evolution of the log radial position of the zone of avoidance for a population of BSS with masses $m_{\rm BSS}=1.5m_*$ (empty circles) and $m_{\rm BSS}=2m_*$ (filled circles), initially placed on randomly chosen, isotropically distributed orbits. Time is measured in units of $10^4$ dynamical crossing times. The orange circles refer to models employing a Holtsmark distribution of force fluctuations, while the blue circles refer to models where a three dimensional Gaussian distribution of fluctuations was used. To guide the eye we plot a local polynomial regression (solid lines) fitted to the $m_{\rm BSS}=2m_*$ zone of avoidance position for Holtsmark (orange) and Gaussian (blue) force kicks. Isotropic orbits are often assumed in most simple models of star clusters (e.g. King, Plummer); in the following we show that our results still hold even in the two extreme anisotropic scenarios where all orbits are circular (see Fig.~\ref{figmin_circul}) or radial (see Fig.~\ref{figmin_radial}).}
    \label{figmin_isotro}
\end{figure}

\begin{figure}
    \centering
    \includegraphics[width=0.9\columnwidth]{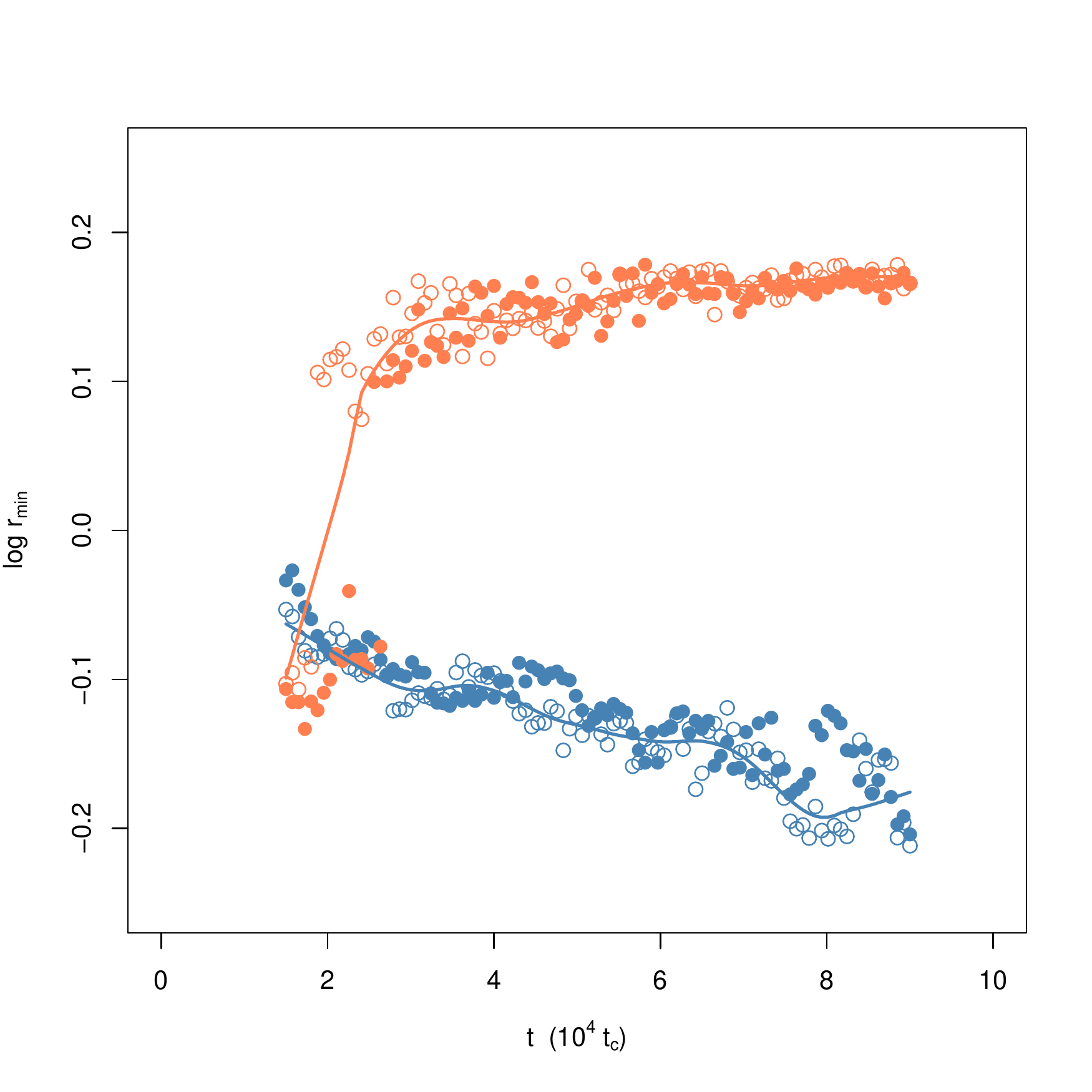}
    \caption{Time evolution of the log radial position of the zone of avoidance for a population of BSS with masses $m_{\rm BSS}=1.5m_*$ (empty circles) and $m_{\rm BSS}=2m_*$ (filled circles), initially placed on purely circular orbits. Time is measured in units of $10^4$ dynamical crossing times. The orange circles refer to models employing a Holtsmark distribution of force fluctuations, while the blue circles refer to models where a three dimensional Gaussian distribution of fluctuations was used. To guide the eye we plot a local polynomial regression (solid lines) fitted to the $m_{\rm BSS}=2m_*$ zone of avoidance position for Holtsmark (orange) and Gaussian (blue) force kicks. Circular- and radial orbits (see Fig.~\ref{figmin_radial}) are two extreme cases that we consider to show that our results are robust to changes in the distribution of BSS orbital angular momenta.}
    \label{figmin_circul}
\end{figure}

\begin{figure}
    \centering
    \includegraphics[width=0.9\columnwidth]{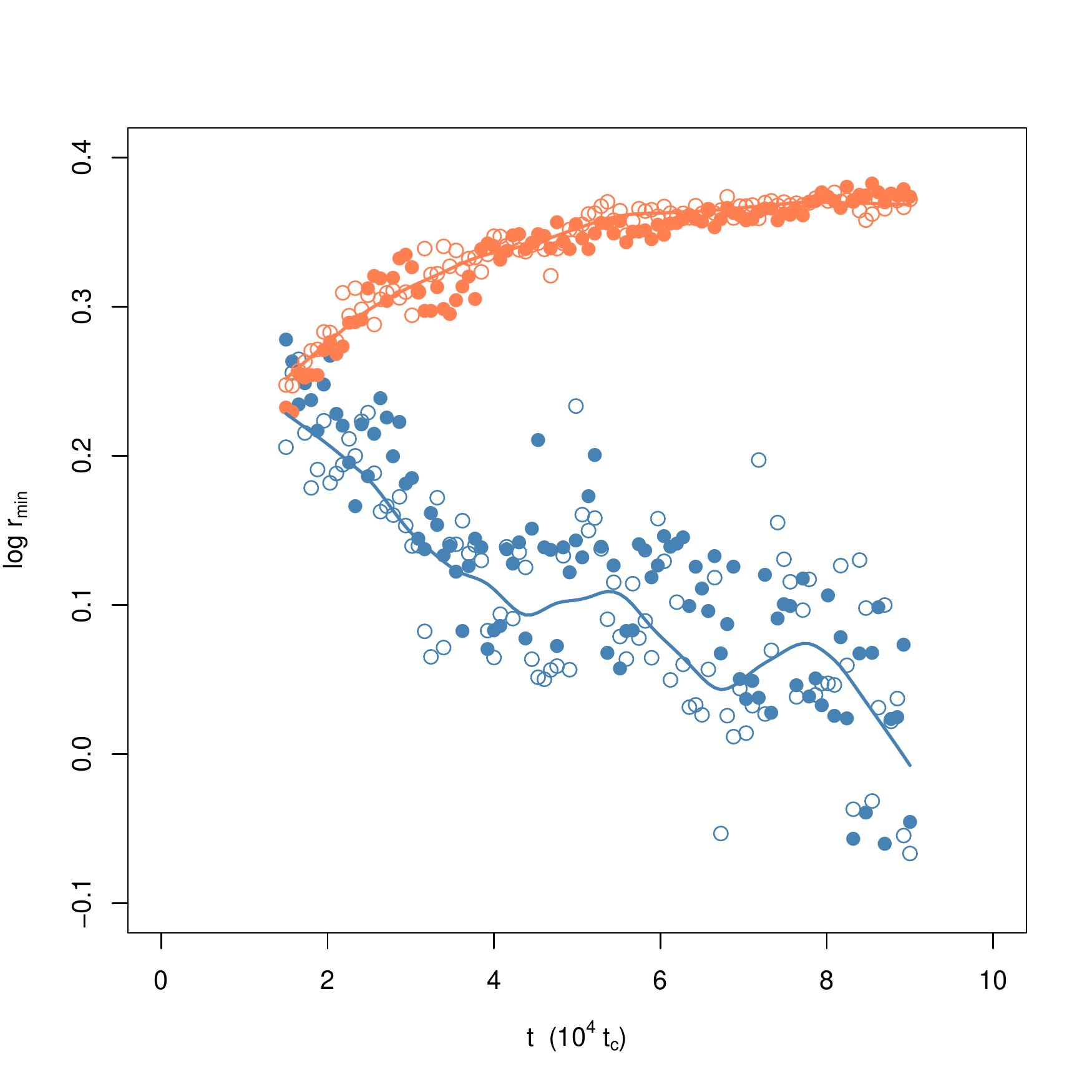}
    \caption{Time evolution of the log radial position of the zone of avoidance for a population of BSS with masses $m_{\rm BSS}=1.5m_*$ (empty circles) and $m_{\rm BSS}=2m_*$ (filled circles), initially placed on purely radial orbits. Time is measured in units of $10^4$ dynamical crossing times. The orange circles refer to models employing a Holtsmark distribution of force fluctuations, while the blue circles refer to models where a three dimensional Gaussian distribution of fluctuations was used. To guide the eye we plot a local polynomial regression (solid lines) fitted to the $m_{\rm BSS}=2m_*$ zone of avoidance position for Holtsmark (orange) and Gaussian (blue) force kicks. Radial- and circular orbits (see Fig.~\ref{figmin_circul}) are two extreme cases that we consider to show that our results are robust to changes in the distribution of BSS orbital angular momenta.}
    \label{figmin_radial}
\end{figure}

In Figure~\ref{holt} and \ref{gauss}, we show the normalized projected radial BSS distribution at increasing dynamical times for a population of tracer particles representing the BSS in simulations using the Holtsmark and the 3d Gaussian distributions of kicks, respectively. In all cases the distribution becomes markedly bimodal after a few hundred dynamical times $t_c$. Note that, in a typical globular cluster a dynamical (crossing) time corresponds to $\approx 10^5$ yr, so the bimodality is established quite rapidly on the cosmological time scale.\\
\indent We show the evolution of the position of the profile's minimum, i.e. the center of the zone of avoidance, as a function of time in \textbf{Fig.~\ref{figmin_isotro}--\ref{figmin_radial}}. Remarkably, using Gaussian kicks fails to reproduce the motion of the minimum towards larger radii over time, while Holtsmark kicks reproduce it correctly as can be seen by comparing the corresponding panels in Fig.~\ref{gauss} and Fig.~\ref{holt}.
We stress the fact that this result holds for all the initial velocity distributions of the BSS explored here, namely isotropic, fully circular and fully radial, so it is not dependent on the specific orbital eccentricity distribution. This was confirmed also by some additional test simulations (not shown here) where the BSS where initialized with different Osipkov-Merritt radially anisotropic profiles.\\
\indent Curiously, regardless of the specific model of force fluctuations, we notice that at large radii the BSS density drops quickly in the case of isotropic or purely radial distribution of orbits, while it takes longer in the case of circular orbits. This behaviour was discussed by \cite{2017MNRAS.471.2537H}, suggesting that it may be due to stars on elongated orbits suffering the effects of dynamical friction much faster than expected based merely on their instantaneous radial position. Our finding confirms this supposition.\\
\indent In addition to the evolution of the minimum, distributions obtained under Gaussian and Holtsmark kicks differ also in the central region, where the former rapidly produce a very strong peak, while the latter show a broader peak that increases slowly over time. In the outskirts, Gaussian kicks result in a slow drop of BSS density, while under Holtsmark kicks the reverse happens and density increases.\\

\indent All these differences find a unified explanation in the fact that Gaussian kicks underestimate the intensity of diffusion, as they do not correctly represent the effects of close encounters, as shown in Fig.~\ref{disth}. Underestimating diffusion allows most stars to fall to the core, forming a central peak and depleting the intermediate regions. As the central peak shrinks over time faster than the distribution in the outskirts gets eroded, the minimum moves towards the inside of the cluster. When two-body kicks are instead correctly accounted for by using the Holtsmark distribution, stars that fall to the core get kicked out, limiting the growth of the central peak and broadening it so that the minimum gets pushed to larger radii. Stars kicked out of the core are responsible also for the rise at large radii observed in the distributions obtained with the Holtsmark kicks. We note that, \cite{1994ApJ...431L.115S} (but see also \citealt{2004ApJ...605L..29M}) in an earlier attempt at modelling the radial BSS distribution, also used a non-Gaussian distribution of Kicks, modelling the effect of triple collisions and inelastic scattering. Remarkably, also in their case the prominent peak at large radii in the relative BSS distribution is recovered.  
\section{Discussion and conclusions}\label{discussion}
\cite{2012Natur.492..393F} have shown that the BSS zone of avoidance evolves in step with the relaxation of the host star cluster by comparing observational data with direct N-body simulations. While simplified, these simulations included a wide range of ingredients and their complex interactions. This is also true for N-body simulations performed later by the same group \cite{2014ApJ...795..169A, 2015ApJ...799...44M} and even more for state-of-the-art Montecarlo simulations that include realistic stellar evolution \cite{2013MNRAS.429.1221H, 2017MNRAS.471.2537H, 2019MNRAS.483.1523S}.\\
\indent The motivation for our work was to do away with this complexity, pinpointing the minimal set of ingredients needed to reproduce the BSS distibution evolution as revealed by observations. We confirm that, in addition to the smooth potential of the host GC, these ingredients are dynamical friction and diffusion as found by \cite{2018ApJ...867..163P}, but in the context of a full three-dimensional model where the stochastic differential equation describing dynamical evolution is solved numerically with diffusion arising from dynamical kicks modeled in a self-consistent way. Additionally, we determined that a correct modelling of these kicks is required to obtain an effective diffusion that reproduces the observed evolution of the radial position of the BSS zone of avoidance, showing how the dynamical clock depends on a delicate equilibrium between diffusion and friction.
This may explain the apparent tension between the results of \cite{2017MNRAS.471.2537H} and \cite{2012Natur.492..393F}\footnote{in addition to the binning choices of the latter discussed by \cite{2017MNRAS.471.2537H}.}: the formation of the zone of avoidance and its correct outwards motion are possible only with the correct recipe for dynamical kicks. If we underestimate kicks (e.g. by assuming they are distributed normally, as shown in this work) not enough diffusion is present to push the BSS minimum outwards; overestimate them and diffusion is too strong and will smooth out the minimum.\\
\indent Except perhaps in a direct N-body model including the correct number of stars (i.e. over $10^6$, which is at the limit of our current technological capabilities) there is in general no guarantee that the distribution of kicks is matched in any simulation setting, in particular in a Monte Carlo.\\
\indent Notwithstanding the large degree of simplification of our model (i.e. assuming the  static and unrealistic Plummer density profile, neglecting stellar evolution and the binary nature of many BSS), the results presented in this work are encouraging and point towards the fact that, at least at first order, the dynamical clock is mainly related to the interplay between dynamical friction and fluctuations of the local gravitational field.\\
\indent A natural follow up of our investigation would be the inclusion of a 'live' star cluster potential accounting for the effects of global dynamical evolution (i.e. core collapse and tidal compression due to the parent galaxy mass distribution) in order to shed some light on how much such collective processes influence the dynamical clock and its effectiveness as a mean to estimate the age of GCs. At the moment, a systematic study using the more realistic King density profiles, allowed to evolve by means of envelope equations for density and potential is underway. In addition, stochastic simulations involving the so-called multiparticle collision technique coupled with standard particle-mesh schemes,  (as recently done in plasma physics, e.g. see \citealt{2017PhRvE..95d3203D,2018CoPP...58..457C}), are underway. By using such methods one is therefore able to compute the collective cluster potential self-consistently with a large number of particles (up to $10^8$) while including the effects of stellar collisions with an operator that preserves locally the kinetic energy and angular and linear momentum of particles. Such methods will allow one to study in more detail mass segregation problems with a larger number of particles than that attainable in direct N-body simulations, with a scheme that is alternative to Monte Carlo at the same computational cost of the simpler Langevin simulations presented in this work.
\begin{acknowledgements}
This project has received funding from the European Union's Horizon $2020$
research and innovation program under the Marie Sk\l{}odowska-Curie grant agreement No. $664931$. One of us (PFDC) wishes to thank the financing from MIUR-PRIN2017 project \textit{Coarse-grained
description for non-equilibrium systems and transport phenomena
(CO-NEST)} n.201798CZL. We wish to thank Prof. Michela Mapelli for discussion and encouragement and the anonymous Referee for his/her important comments that improved the presentations of our results.
\end{acknowledgements}
\bibliographystyle{aa}
\bibliography{ms}

\end{document}